\documentclass[5p]{elsarticle}

\usepackage{amsmath,amsthm,amssymb,amsfonts}
\usepackage[colorlinks,urlcolor=blue,linkcolor=blue,citecolor=blue,anchorcolor=blue]{hyperref}
\usepackage{lineno}
\modulolinenumbers[5]

\begin{document}

\begin{frontmatter}

\title{Vector valley Hall edge solitons in superhoneycomb lattices}

\author[1]{Qian Tang}
\author[2]{Yiqi Zhang\corref{cor}}
\ead{zhangyiqi@xjtu.edu.cn}
\author[3]{Yaroslav V. Kartashov}
\author[2]{Yongdong Li}
\author[4]{Vladimir V. Konotop}
\address[1]{Ministry of Education Key Laboratory for Nonequilibrium Synthesis and Modulation of Condensed Matter, Shaanxi Province Key Laboratory of Quantum Information and Quantum Optoelectronic Devices, School of Physics, Xi'an Jiaotong University, Xi'an 710049, China}
\address[2]{Key Laboratory for Physical Electronics and Devices of the Ministry of Education \& Shaanxi Key Lab of Information
	Photonic Technique, School of Electronic and Information Engineering, Xi'an Jiaotong University, Xi'an 710049, China}
\address[3]{Institute of Spectroscopy, Russian Academy of Sciences, Troitsk, Moscow, 108840, Russia}
\address[4]{Departamento de F\'{\i}sica and Centro de F\'{\i}sica Te\'{o}rica e Computacional, Faculdade de Ci\^{e}ncias, Universidade de Lisboa, Campo Grande, Ed. C8, Lisboa 1749-016, Portugal}
\cortext[cor]{Corresponding author}

\begin{abstract}
Topological edge solitons that bifurcate and inherit topological protection from linear edge states and, therefore, demonstrate immunity to disorder and defects upon propagation, attract considerable attention in a rapidly growing field of topological photonics. Valley Hall systems are especially interesting from the point of view of realization of topological edge solitons because they do not require external or artificial magnetic fields or longitudinal modulations of the underlying potential for the emergence of the topological phases. Here we report on the diverse types of vector valley Hall edge solitons forming at the domain walls between superhoneycomb lattices, including bright-dipole, bright-tripole, dark-bright, and dark-dipole solitons. In contrast to conventional scalar topological solitons, such vector states can be constructed as envelope solitons on the edge states from different branches and with different Bloch momenta. Such vector solitons can be remarkably robust, they show stable long-distance propagation and can bypass sharp bends of the domain wall. The existence of the counter-propagating valley Hall edge solitons at the same domain wall allows us to study their structural robustness upon collisions that can be nearly elastic. Our results illustrate richness of soliton families in the valley Hall systems and open new prospects for the light field manipulation and design of the nonlinear topological functional devices.
\end{abstract}

\begin{keyword}
valley Hall effect \sep Floquet topological insulator \sep Topological edge soliton
\end{keyword}

\end{frontmatter}


\section{Introduction}

During the last decade, the nonlinear topological photonics \cite{smirnova.apr.7.021306.2020} has grown up into a significant branch of the topological photonics \cite{lu.np.8.821.2014,ozawa.rmp.91.015006.2019,kim.lsa.9.130.2020,ota.nano.9.547.2020, leykam.nano.9.4473.2020,segev.nano.10.425.2021,parto.nano.10.403.2021,wang.fo.13.50.2020,liu.col.19.052602.2021}. Nonlinearity in topological systems enables novel interesting phenomena that are not available in purely linear systems, and significantly enriches tools for control of transport, localization, and internal structure of excitations. Typical examples of such nonlinear phenomena include nonlinearity-induced topological transitions \cite{maczewsky.science.370.701.2020},
nontrivial coupling of light into topological system \cite{xia.light.9.147.2020}, bistability effects in pumped dissipative systems \cite{kartashov.prl.119.253904.2017,zhang.lpr.13.1900198.2019}, lasing in topological edge states \cite{harari.science.359.eaar4003.2018,bandres.science.359.eaar4005.2018,dikopoltsev.science.373.1514.2021,bahari.science.358.636.2017,kartashov.prl.122.083902.2019,ivanov.apl.4.126101.2019,zeng.nature.578.246.2020,zhong.lpr.14.2000001.2020,gong.acs.7.2089.2020,zhong.apl.6.040802.2021,choi.nc.12.3434.2021}, and topological edge solitons \cite{ablowitz.pra.90.023813.2014, ablowitz.pra.96.043868.2017,ablowitz.ol.40.4635.2015,leykam.prl.117.143901.2016,ivanov.ol.45.1459.2020,ivanov.ol.45.2271.2020,ivanov.acs.7.735.2020,ivanov.pra.103.053507.2021,ivanov.lpr.202100398,zhong.ap.3.056001.2021,tang.oe.29.39755.2021,ren.nano.10.3559.2021,tang.rrp.74.405.2022}. Topological edge solitons are specific states that bifurcate from linear topological edge states. They have propagation constant belonging to topological forbidden gap. On this reason they benefit from topological protection in the presence of defects or disorder. Nonlinearity provides localization of such states along the edge of the insulator in addition to localization across the edge due to their topological nature, allowing such solitons to travel along the edge without broadening. Topological edge soliton has been reported in Floquet systems \cite{ablowitz.pra.90.023813.2014, ablowitz.pra.96.043868.2017, ablowitz.ol.40.4635.2015, leykam.prl.117.143901.2016, ivanov.ol.45.1459.2020, ivanov.ol.45.2271.2020, ivanov.acs.7.735.2020, ivanov.pra.103.053507.2021, ivanov.lpr.202100398, mukherjee.science.368.856.2020, mukherjee.prx.11.041057.2021}, in polariton microcavities \cite{kartashov.optica.3.1228.2016, gulevich.sr.7.1780.2017, li.prb.97.081103.2018, zhang.pra.99.053836.2019}, see also first realization of polariton topological insulator \cite{klembt.nature.562.552.2018}, and in valley Hall systems \cite{zhong.ap.3.056001.2021, tang.oe.29.39755.2021, ren.nano.10.3559.2021,tang.rrp.74.405.2022}. In addition, topological solitons bifurcating from corner states in higher-order topological insulators \cite{kirsch.np.17.995.2021,hu.light.10.164.2021}, interface states in truncated dimer and trimer topological chains \cite{meier.nc.7.13986.2016, xia.light.9.147.2020, guo.ol.45.6466.2020, ma.pre.104.054206.2021, kartashov.prl.128.093901.2022}, and edge solitons in truncated lattices induced in atomic vapors \cite{zhang.nc.11.1902.2020} have been reported.

Among all these topological systems, the systems relying on the valley Hall effect \cite{xue.apr.2.2100013.2021,tang.lpr.16.2100300.2022} are very promising for experimental realization of topological edge solitons, since they do not require sophisticated longitudinal modulations of the refractive index landscapes leading to additional losses as in Floquet systems \cite{ablowitz.pra.90.023813.2014, ablowitz.pra.96.043868.2017, ablowitz.ol.40.4635.2015, leykam.prl.117.143901.2016, ivanov.ol.45.1459.2020, ivanov.ol.45.2271.2020, ivanov.acs.7.735.2020, ivanov.pra.103.053507.2021, ivanov.lpr.202100398,fu.prl.128.154101.2022} or external or artificial magnetic fields as in structured polariton microcavities \cite{kartashov.optica.3.1228.2016, gulevich.sr.7.1780.2017, li.prb.97.081103.2018, zhang.pra.99.053836.2019}. In valley Hall systems edge states emerge on the domain wall between two lattice regions due to breaking of the inversion symmetry of the lattice. This symmetry breaking can be achieved, for example, by introducing different detuning into depths of lattice channels at both sides of the domain wall \cite{xue.apr.2.2100013.2021,tang.lpr.16.2100300.2022}. So far, only scalar bright and dark \cite{zhong.ap.3.056001.2021, tang.oe.29.39755.2021, ren.nano.10.3559.2021, smirnova.prr.3.043027.2021} or specific Dirac \cite{smirnova.lpr.13.1900223.2019} edge solitons were encountered in valley Hall photonic systems. However, such systems may also offer unique conditions for realization of vector edge solitons, characterized by more than one components that simultaneously make their construction more challenging. Among the requirements for the existence of such solitons is the presence of two (or more) edge states with equal group velocities at a given domain wall.

While domain walls between two non-strained detuned honeycomb lattices possessing $C_3$ symmetry do not support more than one edge state with the same group velocity \cite{tang.oe.29.39755.2021, ren.nano.10.3559.2021}, such states may form at the domain walls between strained honeycomb lattices \cite{xi.pjr.8.b1.2020} (such lattices are characterized by large valley Chern numbers and support more than one valley Hall edge state in a given gap according to the bulk-edge correspondence principle \cite{ozawa.rmp.91.015006.2019}), or at the domain walls between so-called superhoneycomb (or edge-centered honeycomb) lattices \cite{zhong.adp.529.1600258.2017,lan.prb.85.155451.2012}. Spectra of lattices of the latter type possess two sets of pseudospin-1/2 Dirac cones between first two and last two bands in the 5-band spectrum. Thus, domain wall between two such detuned lattices may support two different coexisting valley Hall edge states, whose group velocities can coincide. In addition, such superhoneycomb lattices still possess $C_3$ symmetry allowing design of the domain walls with sharp corners.

In this work, we introduce a rich variety of vector valley Hall edge solitons available on the domain wall between two superhoneycomb lattices with broken inversion symmetry due to properly adjusted refractive indices of different waveguides. We report on the existence of bright-dipole, bright-tripole, dark-bright, and dark-dipole vector valley Hall edge solitons. Such solitons exist as exceptionally robust objects maintaining their internal structure upon collisions and passage through sharp bends of the domain wall.

\section{Band structures and valley Hall edge states}

\begin{figure}[htbp]
	\centering
	\includegraphics[width=\columnwidth]{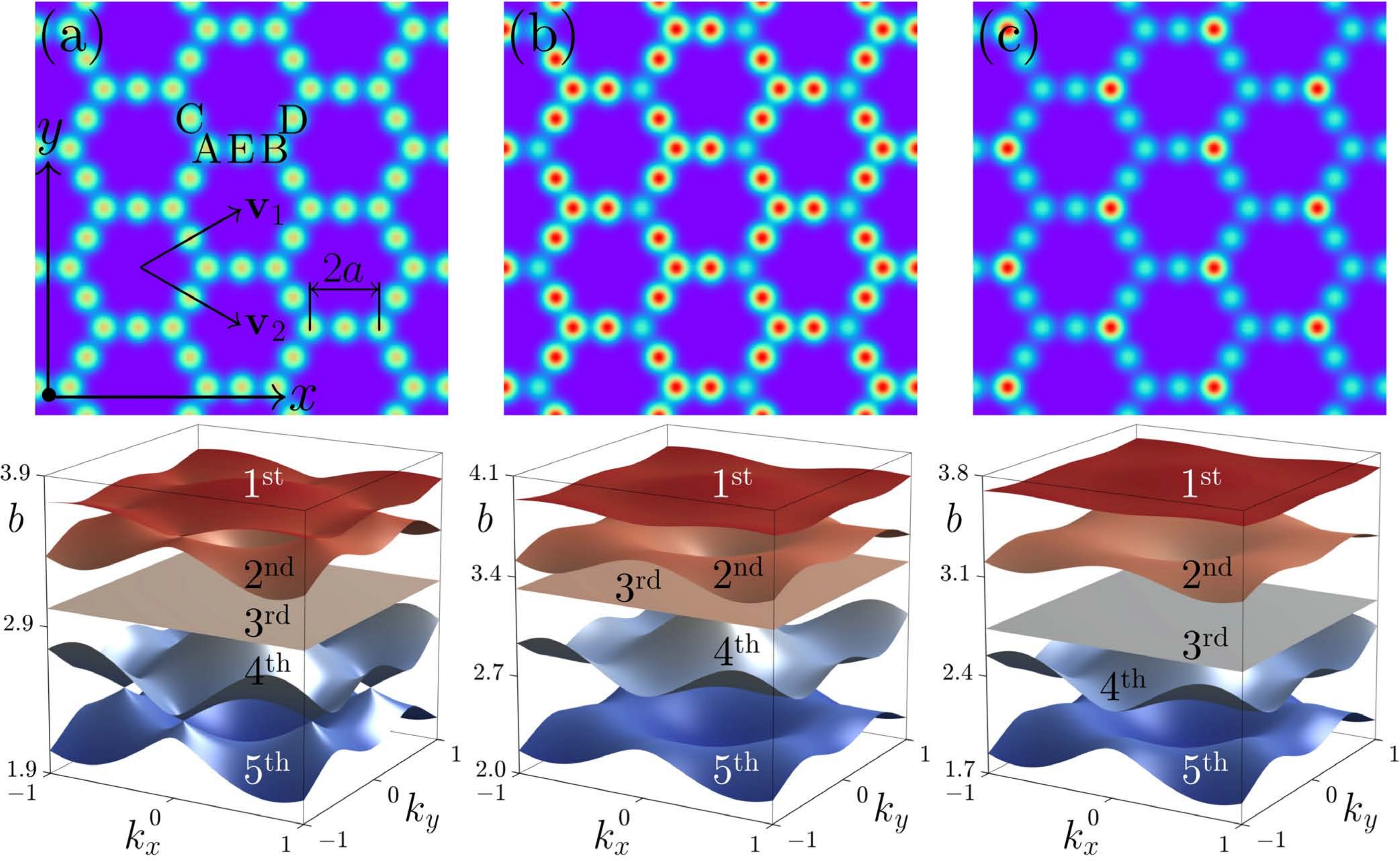}
	\caption{Superhoneycomb photonic lattices with identical or detuned sites (first row) and corresponding band structures (second row). Panel (a) shows the lattice with inversion symmetry, where all sites have the same depth. There are 5 sites (labeled as A, B, C, D and E) in one unit cell with ${\bf v}_1=[3a,\sqrt{3} a]$ and ${\bf v}_2=[3a,-\sqrt{3} a]$ being primitive lattice vectors. The inter-site spacing is $a=1.4$ and the depth of this lattice is $p=9.5$. (b) Inversion-symmetry-broken superhoneycomb lattice with the depth of sites B being $p-\delta$ and depths of all other sites being $p+\delta$, with detuning $\delta=0.55$. (c) Similar inversion-symmetry-broken lattice, where sites B have depth $p+\delta$ and all other sites have depth $p-\delta$, with $\delta=0.55$. Here and below we set $d=0.5$.}
	\label{fig1}
\end{figure}

Propagation of a light beam in a medium with focusing Kerr nonlinearity and imprinted shallow lattice is governed by the continuous dimensionless nonlinear Schr\"{o}dinger equation
\begin{equation}\label{eq1}
	i \frac{\partial \psi}{\partial z}=
	-\frac{1}{2} \left(\frac{\partial^{2}}{\partial x^{2}}+\frac{\partial^{2}}{\partial y^{2}}\right) \psi - \mathcal{R}(x, y) \psi  -|\psi|^{2} \psi.
\end{equation}
Here $\psi$ is a complex-valued envelope of the electric field, while lattice potential is described by the function
\begin{equation}\label{eq2}
	\mathcal{R}(x, y)=p\sum_{n, m} \exp\left(-\frac{(x-x_n)^2 + (y-y_m)^2}{d^2}\right)
\end{equation}
where $d$ is the width of individual lattice channels (sites) and $(x_n, y_m)$ are the positions of sites with index $n,m$ in a superhoneycomb grid. The profile of such superhoneycomb lattice is presented in Fig. \ref{fig1}(a). The primitive lattice vectors are ${\bf v}_1=[3a,\sqrt{3} a]$ and ${\bf v}_2=[3a,-\sqrt{3} a]$, where $a$ is the distance between two nearest-neighbor channels. The superhoneycomb lattice has 5 sites in one unit cell that are labeled as A, B, C, D and E in Fig. \ref{fig1}(a). If one removes sites C, D and E, the superhoneycomb lattice transforms into a simple honeycomb lattice.

In the top row of Fig. \ref{fig1} we show examples of superhoneycomb lattices. Corresponding two-dimensional band structures are shown in the bottom row. In the lattice shown in Fig.~\ref{fig1}(a), all sites have the same depth $p=9.5$. In the inversion-symmetry-broken lattice from Figs.~\ref{fig1}(b) the depth of sites B, $p_{\rm B}=p-\delta$, is smaller than depths of all other sites, $p_{\rm A,C,D,E}=p+\delta$, where $\delta$ is the refractive index detuning. In the lattice from Fig. \ref{fig1}(c) the depth of sites B, $p_{\rm B}=p+\delta$, is instead larger than depths of all other sites $p_{\rm A,C,D,E}=p-\delta$. Generally speaking, since superhoneycomb lattice has five sites in the unit cell, its inversion symmetry can be broken in several different ways, beyond the one that we use here. At the same time, Dirac cones visible in Fig.~\ref{fig1}(a) between the 1st and 2nd bands and between 4th and 5th bands, which are crucial for the appearance of the edge states, are affected mostly by modulations of sites A and B, while detuning of sites C, D and E does not lead to desirable transformations of the band structure. In the bottom row of Fig. \ref{fig1} one can see how introduction of detuning leads to simultaneous opening of two gaps between former Dirac points.

\begin{figure}[htbp]
	\centering
	\includegraphics[width=\columnwidth]{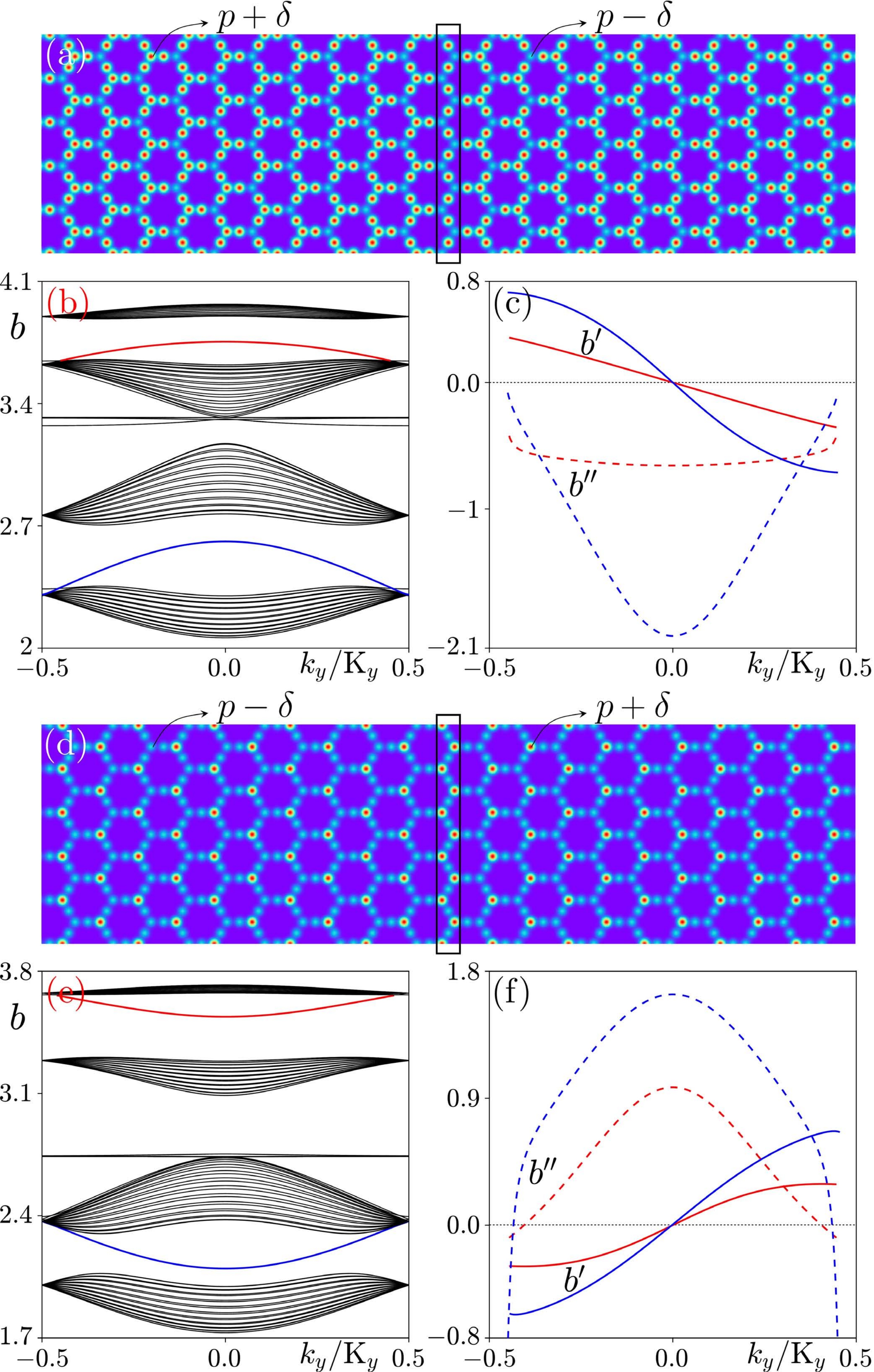}
	\caption{(a) A superhoneycomb lattice composed of the lattice shown  in Fig. \ref{fig1}(b) and its conjugate, with the domain wall highlighted by the rectangle. (b) The corresponding projected band structure. Red and blue curves are the valley Hall edge states, while the black curves correspond to the bulk states.  (c) First-order (solid curves) and second-order (dashed curves) derivatives ($b'$ and $b''$) of the valley Hall edge states shown in (b). (d,e,f) Setup is as (a,b,c), but the composited superhoneycomb lattice composed of lattice in Fig. \ref{fig1}(c) and its conjugate.}
	\label{fig2}
\end{figure}

Next we define ``conjugate'' lattices to the structures depicted in Fig. \ref{fig1}(b),(c). For example, the lattice with $p_{\rm A}=p-\delta$ and $p_{\rm B,C,D,E}=p + \delta$ is the conjugate one to the lattice from Fig. \ref{fig1}(b). By combining the original superhoneycomb lattice and its conjugate counterpart, we created two different types of interfaces or zigzag-zigzag domain walls depicted in Figs.~\ref{fig2}(a) and (d) (domain walls are highlighted by the rectangles). Such composite lattice structures are periodic along the $y-$direction, $\mathcal{R}(x, y) = \mathcal{R}(x, y + {\rm Y})$, with the period ${\rm Y} = 2\sqrt{3}a$, and are constrained in the $x$ direction [to ensure that states at the domain wall are not affected by outer lattice edges, we consider very wide structures in the $x$-direction extending far beyond windows depicted in Figs.~\ref{fig2}(a) and (d)].

In the linear limit the stationary eigenvalue problem corresponding to Eq.~(\ref{eq1}), can be written as
\begin{equation}\label{eq3}
	b(k_y) u = \frac{1}{2} \left( \frac{\partial^2}{\partial x^2} + \frac{\partial^2}{\partial y^2}
	+2 i k_{y} \frac{\partial}{\partial y} - k_y^2 \right) u + {\mathcal R} u,
\end{equation}
where we used the Bloch ansatz
$
\psi = \phi(x,y) \exp[i b(k_y) z]
$,
where  $\phi(x,y)=u(x,y)\exp(ik_y y)$ with $u(x,y)=u(x,y+{\rm Y})$, and $k_y \in [-{\rm K}_y/2,{\rm K}_y/2)$ is the Bloch momentum in the first Brillouin zone of width ${\rm K}_y=2\pi/{\rm Y}$. Since we consider large, but finite in $x$ lattices, the function $u(x,y) \to 0$ when $x \to \pm \ell/2$, where $\ell \gg a$ is the size of integration window in $x$. Equation (\ref{eq3}) was solved numerically using plane-wave expansion method. The so-obtained ``projected'' band structure is depicted in Fig.~\ref{fig2}(b) for the structure produced by combination of lattice from Fig. \ref{fig1}(b) and its conjugate, where the red and blue curves correspond to the valley Hall edge states and black curves correspond to the bulk states. Both types of the valley Hall edge states (shown by the red and blue lines) are localized at the same domain wall, thus creating the necessary conditions for bifurcation of the families of vector valley Hall edge solitons. The first-order, $b'=db/dk_y$, and second-order, $b''=d^2b/dk_y^2$, derivatives of the propagation constants of the valley Hall edge states are shown in Fig. \ref{fig2}(c). For this type of the domain wall $b''<0$ for both edge states, as required for the construction of bright solitons in the focusing Kerr medium. If the domain wall is created by the superhoneycomb lattice from Fig. \ref{fig1}(c) and its conjugate [Fig. \ref{fig2}(d)], one also obtains two valley Hall edge states, shown by the red and blue lines in Fig.~\ref{fig2}(e). However, the dispersion coefficients for these edge states depicted in Fig. \ref{fig2}(f) are now positive $b''>0$ nearly in the entire first Brillouin zone. Thus, the domain wall from Fig.~\ref{fig2}(d) may support dark solitons. These results confirm that superhoneycomb lattices may serve as an excellent platform for the construction of various vector solitons and it offers conditions for the existence of such solitons that are hard to meet for other lattice types.

Superhoneycomb lattices considered here can be realized experimentally using well-established method of direct fs-laser writing in transparent dielectrics, such as fused silica \cite{rechtsman.nature.496.196.2013, stuetzer.nature.560.461.2018, mukherjee.science.368.856.2020, maczewsky.science.370.701.2020, kirsch.np.17.995.2021, tan.ap.3.024002.2021}. Assuming wavelength of laser radiation of $800~\rm nm$ and introducing the characteristic transverse scale of $10~\mu \rm m$ corresponding to dimensionless values $x,y=1$, one obtains that propagation distance $z=1$ corresponds to $1.14 ~\textrm{mm}$, while lattice considered above has separation of $14~\mu \rm m$ between neighbouring waveguides, $5~\mu \rm m$ widths of the individual waveguides, and depth $p\sim9.5$ corresponding to real refractive index modulation $10^{-3}$.

\section{Envelopes of vector solitons}

To construct vector topological edge solitons we consider bifurcation of the nonlinear families of solutions from the linear Bloch states~\cite{ivanov.acs.7.735.2020, ivanov.pra.103.053507.2021} characterized by equal group velocities $b_\alpha^{\prime}=b_\beta^{\prime}=:-v$ (here $\alpha,\beta$ are the indices of the edge states on which we construct solitons). To describe their envelopes we introduce the following ansatz:
\begin{equation}\label{eq4}
	\begin{split}
	\psi = & A_\alpha(Y,z) \phi_{\alpha,k_{y,\alpha}} \exp(ib_{\alpha,k_{y,\alpha}}z) + \\
	& A_\beta(Y,z) \phi_{\beta,k_{y,\beta}} \exp(ib_{\beta,k_{y,\beta}}z) ,
	\end{split}
\end{equation}
where $A_{\alpha,\beta}$ are the slowly-varying envelopes and $Y = y - v z$ is the coordinate in the reference frame moving with velocity $v$. The evolution of the envelopes is then governed by the coupled nonlinear Schr\"odinger equations that can be derived using a variant of the multiple-scale expansion procedure described in \cite{ivanov.acs.7.735.2020}:
\begin{equation}\label{eq5}
	\begin{split}
		i\frac{\partial A_\alpha}{\partial z} &=  \frac{b_\alpha^{\prime\prime}}{2} \frac{\partial^2 A_\alpha}{\partial Y^2} - \left( \chi_\alpha |A_\alpha|^2 + 2\chi |A_\beta|^2 \right) A_\alpha, \\
		i\frac{\partial A_\beta}{\partial z}  &=  \frac{b_\beta^{\prime\prime}}{2}  \frac{\partial^2 A_\beta}{\partial Y^2} - \left( \chi_\beta |A_\beta|^2 + 2\chi |A_\alpha|^2 \right) A_\beta,
	\end{split}
\end{equation}
where $\chi = \int_S |\phi_\alpha|^2|\phi_\beta|^2 dxdy$ and $\chi_{\nu} = \int_S |\phi_{\nu}|^4 dxdy$ with $\nu=\alpha,\beta$ are the effective nonlinearity coefficients, where Bloch waves are normalized as $ \int_S |\phi_\nu|^2 dxdy=1$, $S$ is the entire area of the array. Solutions of Eqs. (\ref{eq5}) can be written in the form $A_\nu(Y,z)=w_\nu(Y)\exp(ib_\nu^{\rm nl}z)$, where $b_\nu^{\rm nl}$ is the nonlinearity-induced propagation constant shift, which should be sufficiently small to guarantee that the profile $w_\nu(Y)$ is broad and fulfills the requirement of slowly varying envelope. For dispersion coefficients of the same sign, the system (\ref{eq5}) admits the simplest possible solutions with similar functional profiles (like bright-bright solitons for negative $b_\alpha^{\prime\prime},b_\beta^{\prime\prime}$ or dark-dark ones for positive $b_\alpha^{\prime\prime},b_\beta^{\prime\prime}$). Being the simplest possible solutions of the system, such states are stable and show behaviour similar to that of scalar solitons. Therefore, further we will be interested in construction of solutions with different functional profiles (different symmetries) in two components. The profiles $w_\nu(Y)$ of such solutions for the parameters of our model can be obtained numerically from Eqs. (\ref{eq5}) using Newton method.

\begin{figure}[htbp]
	\centering
	\includegraphics[width=\columnwidth]{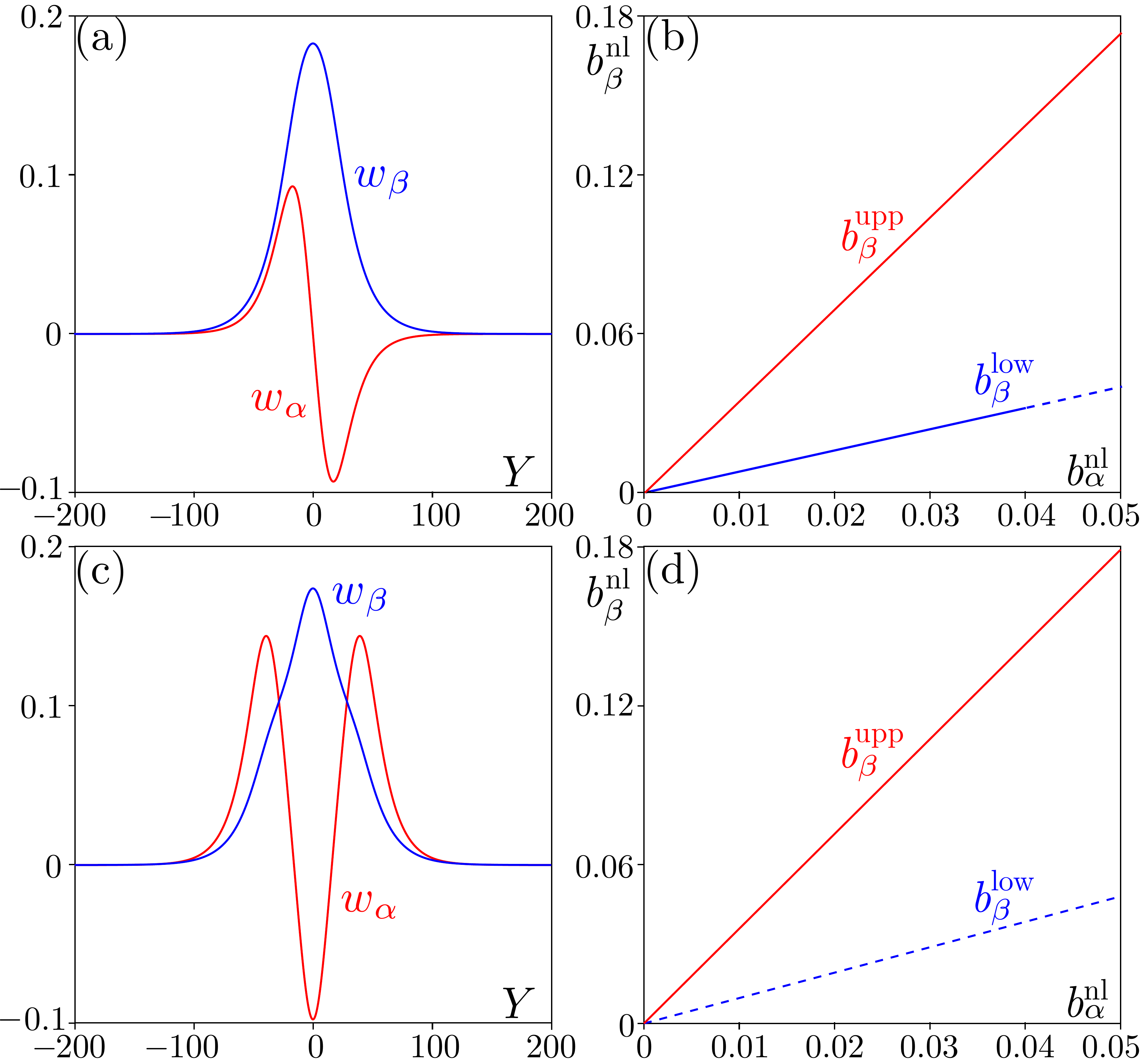}
	\caption{(a) Envelopes of the components of bright-dipole vector soliton. (b) Existence domain $b_\beta^{\rm low}<b_\beta^{\rm nl}<b_\beta^{\rm upp}$ of the bright-dipole vector soliton obtained for different $b_\alpha^{\rm nl}$. (c,d) Envelopes and existence domain for bright-tripole vector soliton. Parameters for the component based on the red valley Hall edge state from Figs. \ref{fig2}(b,c): $k_{y,\alpha}=0.31{\rm K}_y$, $b_\alpha^{\rm nl}=0.0014$ and $b_\alpha^{\prime\prime}=-0.6013$. Parameters for the component based on the blue valley Hall edge state from Figs. \ref{fig2}(b,c): $k_{y,\beta}=0.102{\rm K}_y$, $b_\beta^{\rm nl}=0.003$ and $b_\beta^{\prime\prime}=-1.8188$. The dashed line in (b,d) implies that the envelope is unstable at a given border, while solid lines mean that the envelope is stable. Effective nonlinear coefficients for this choice of $k_{y,\alpha},k_{y,\beta}$ are $\chi_\alpha=0.1061$, $\chi_\beta=0.1369$ and $\chi=0.0637$.}
	\label{fig3}
\end{figure}

First, we construct the envelopes for edge solitons on the domain wall depicted in Fig. \ref{fig2}(a) that supports two valley Hall edge states with negative second-order derivatives. For their construction we choose two edge states from Fig. \ref{fig2}(b) from the red branch at $k_{y,\alpha}=0.31{\rm K}_y$ and blue branch at $k_{y,\beta}=0.102 {\rm K}_y$ with equal group velocities $b_\alpha^{\prime},b_\beta^{\prime}$. Corresponding dispersion and effective nonlinear coefficients are given in the caption to Figs. \ref{fig3}. The examples of numerically obtained from Eqs. (\ref{eq5}) envelopes of bright-dipole and bright-tripole vector solitons are shown in Figs. \ref{fig3}(a) and \ref{fig3}(c), respectively. For a fixed nonlinear shift of propagation constant $b_\alpha^{\rm nl}$ such solitons exist within limited interval of $b_\beta^{\rm nl}$ values, $b_\beta^{\rm low} \le b_\beta^{\rm nl} \le b_\beta^{\rm upp}$, that nearly linearly expands with increase of $b_\alpha^{\rm nl}$. The domains of existence for bright-dipole and bright-tripole solitons are shown in Figs. \ref{fig3}(b) and \ref{fig3}(d), respectively. Notice that in the frames of envelope equations (\ref{eq5}) these domains can be formally continued to much larger values of $b_\alpha^{\rm nl}$ and $b_\beta^{\rm nl}$, where the borders of the domain may depart from linear functions, but in reality the considered domain should be consistent with the assumption that envelope soliton is broad and covers many $y$-periods of the array. On the upper edge of the existence domain $\alpha$ component disappears and soliton transforms into scalar state with only $\beta$ component being nonzero. Close to the lower edge of the existence domain solitons transform into several well-separated bright-bright states, whose $\beta$ components gradually vanish. We have also checked stability of such envelope solitons by adding a small-scale perturbation into input field distributions with an amplitude up to $10\%$ of soliton's amplitude and propagating them over large distance. In Figs. \ref{fig3}(b) and \ref{fig3}(d), the unstable solutions are indicated by the dashed lines, while solid lines correspond to stable states. All such vector states are stable close to the upper border of the existence domain, where bell-shaped $\beta$ component dominates.

\begin{figure}[htbp]
	\centering
	\includegraphics[width=\columnwidth]{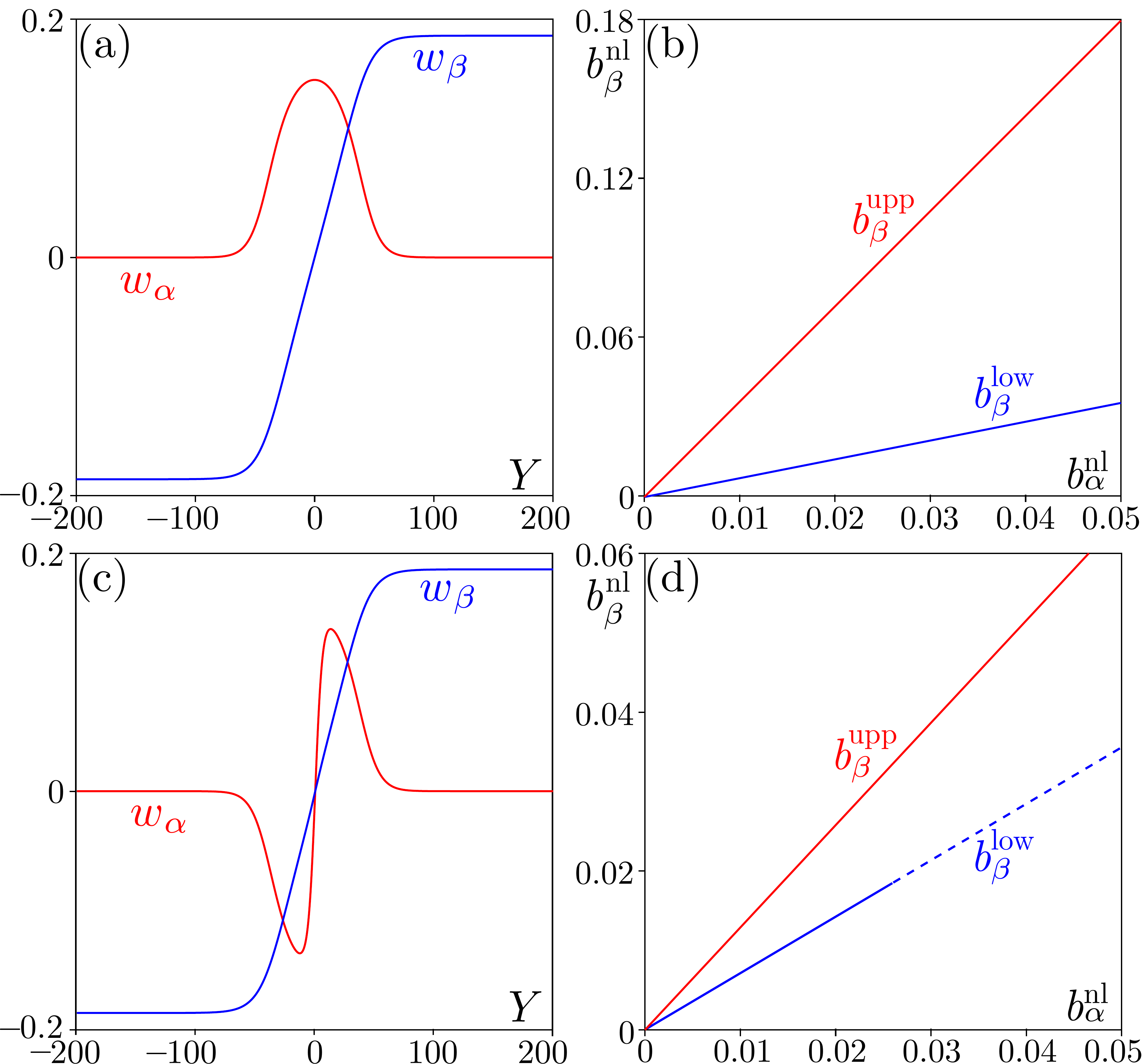}
	\caption{(a,b) Example of envelopes and existence domain on the $b_\alpha^{\rm nl},b_\beta^{\rm nl}$ plane for dark-bright vector soliton. (c,d) Example of envelopes and existence domain on the $b_\alpha^{\rm nl},b_\beta^{\rm nl}$ plane for dark-dipole vector soliton. Parameters of the component based on the red valley Hall edge state from Figs. \ref{fig2}(e,f): $k_{y,\alpha}=0.344{\rm K}_y$, $b_\alpha^{\rm nl}=0.0045$ and $b_\alpha^{\prime\prime}=0.1551$. Parameters of the component based on the blue valley Hall edge state from Figs. \ref{fig2}(e,f): $k_{y,\beta}=0.14{\rm K}_y$, $b_\beta^{\rm nl}=0.0035$ and $b_\beta^{\prime\prime}=1.458$. Effective nonlinear coefficients are $\chi_\alpha=0.2006$, $\chi_\beta=0.1010$ and $\chi=0.0865$.}
	\label{fig4}
\end{figure}

Second, we construct the envelopes for the case of the domain wall depicted in Fig. \ref{fig2}(d). Now we use edge states from Fig. \ref{fig2}(e) from the red branch at $k_{y,\alpha}=0.344{\rm K}_y$ and blue branch at $k_{y,\beta}=0.14{\rm K}_y$ with equal group velocities $b_\alpha^{\prime},b_\beta^{\prime}$ and positive second-order derivatives. Associated dispersion and effective nonlinear coefficients are given in the caption to Fig. \ref{fig4}. Examples of the dark-bright and dark-dipole vectors solitons numerically obtained from Eqs. (\ref{eq5}) are shown in Figs. \ref{fig4}(a) and \ref{fig4}(c), respectively. Such solitons exist for positive $b_\alpha^{\rm nl}, b_\beta^{\rm nl}$ values, similarly to solitons from Fig. \ref{fig3}. Even though $\beta$ component in such states has nonzero asymptotics, typical for dark solitons, the other $\alpha$ component features localized shape. For such states the width of the existence domain in $b_\beta^{\rm nl}$ also increases with increase of $b_\alpha^{\rm nl}$, see Figs. \ref{fig4}(b) and \ref{fig4}(d). Similarly to localized states discussed in Fig. \ref{fig3}, $\alpha$ component disappears close to the upper edge $b_\beta^{\rm upp}$ of the existence domain and solitons transform into scalar dark states. In contrast, when $b_\beta^{\rm nl}$ approaches lower edge of the existence domain, $\alpha$ component substantially broadens and develops flat-top shape expelling $\beta$ component from the spatial region, where $\alpha$ component is nonzero, so that soliton state effectively delocalizes at $b_\beta^{\rm nl}=b_\beta^{\rm low}$ . Dark-bright vector states obtained here also can be stable (in particular, they are always stable close to the upper border of the existence domain), as indicated by solid lines at the borders of the existence domain.

\section{Propagation dynamics of the valley Hall edge solitons}

To demonstrate the formation of topological edge solitons in full two-dimensional (2D) valley Hall system and to confirm that envelope equations (\ref{eq5}) indeed accurately describe envelopes of such solitons, we constructed input 2D field distribution in accordance with Eq. (\ref{eq4}), by superimposing the envelopes $A_\nu(Y,z=0)=w_\nu(Y)$ obtained from (\ref{eq5}) on corresponding linear valley Hall edge states $\phi_{\nu,k_{y,\nu}}$. The so-constructed initial distribution $\psi(x,y,z=0)$ that simultaneously contains contributions from two edge states with different Bloch momenta $k_{y,\nu}$ (reflecting its vector nature) was propagated using full 2D Eq. (\ref{eq1}). Notice that we work here with sufficiently broad states covering many array periods to minimize higher-order dispersion and to ensure validity of the envelope equations, hence diffraction length for such states is determined by the width of their envelopes and to make it obvious we propagate over sufficiently large distances. We found that in all cases the so-constructed initial state maintains its initial profile over huge propagation distances, even though it moves along the domain wall because it is constructed on the edge states with nonzero group velocity. This fully confirms the accuracy of our approach, as illustrated below.

\begin{figure}[htbp]
	\centering
	\includegraphics[width=\columnwidth]{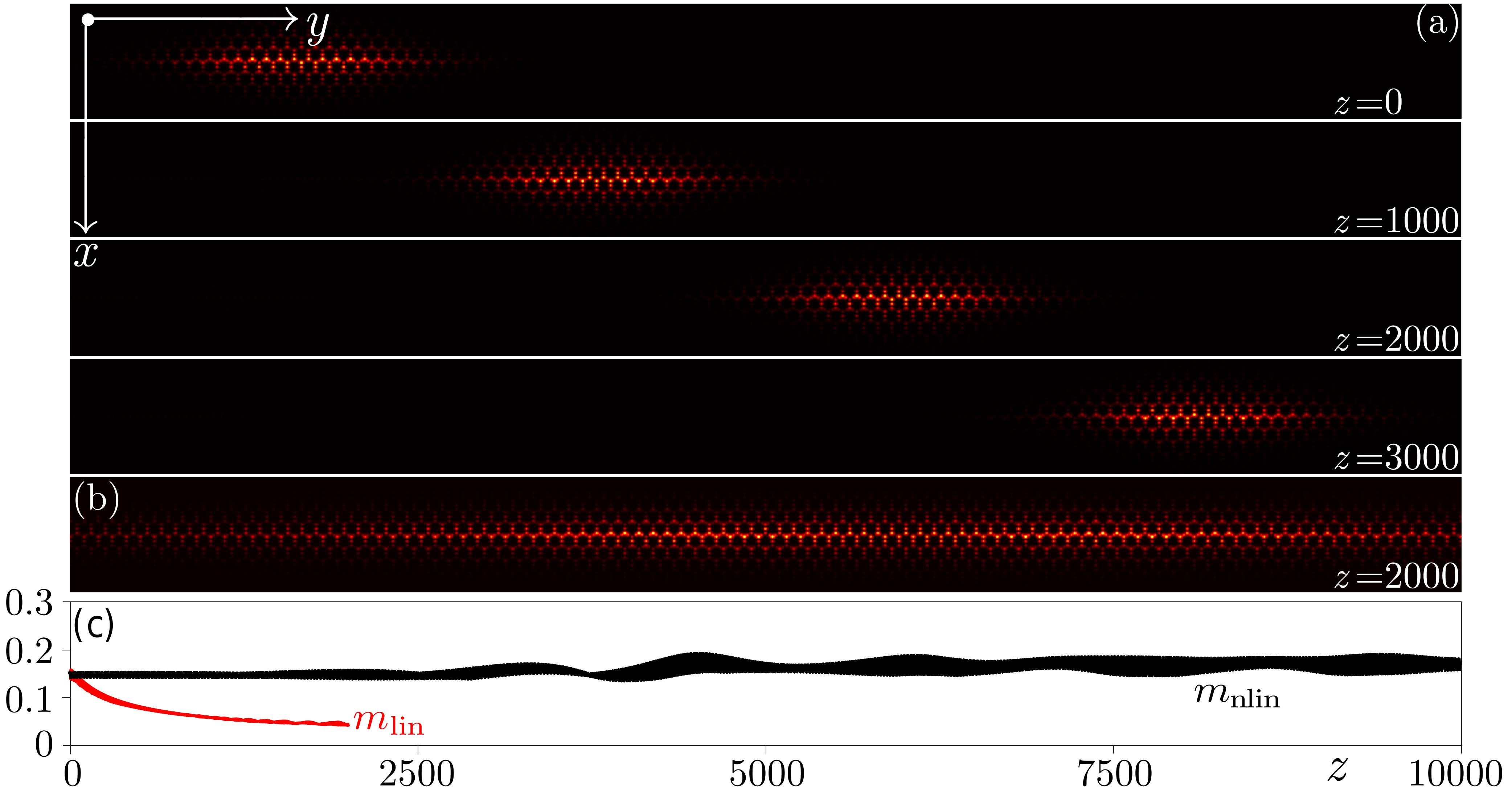}
	\caption{Propagation dynamics of the bright-dipole vector soliton. (a) Selected field modulus distributions for soliton at different propagation distances. (b) Diffraction of the same input state as in (a) at $z=2000$ in the absence of nonlinear term in Eq. (\ref{eq1}). (c) Peak amplitude of the beam during nonlinear ($m_{\rm nlin}$) and linear ($m_{\rm lin}$) propagation. Panels in (a) and (b) are shown in the window $-20\le x\le20$ and $-242.5\le y\le242.5$. Parameters of the component based on the red valley Hall edge state in Figs. \ref{fig2}(b,c): $k_{y,\alpha}=0.122{\rm K}_y$, $b_\alpha^{\rm nl} = 0.002$ and $b_\alpha^{\prime\prime}=-0.6486$. Parameters of the component based on the blue valley Hall edge state in Figs. \ref{fig2}(b,c): $k_{y,\beta}=0.04{\rm K}_y$, $b_\beta^{\rm nl}=0.004$ and $b_\beta^{\prime\prime}=-1.9742$. The velocity of soliton is $v\sim 0.1033$.}
	\label{fig5}
\end{figure}

\begin{figure}[htbp]
	\centering
	\includegraphics[width=\columnwidth]{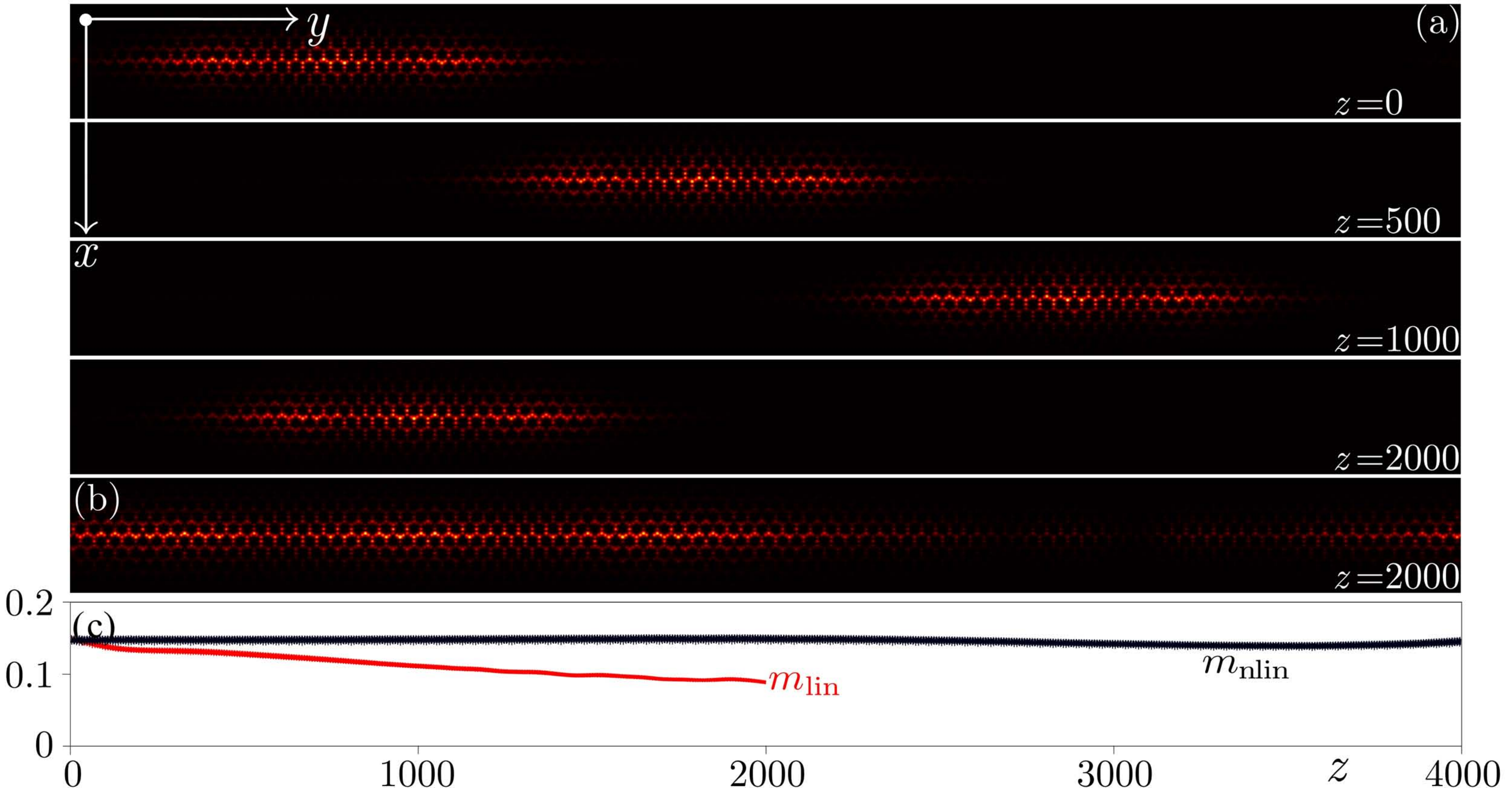}
	\caption{Setup is as Fig. \ref{fig5}, but for the bright-tripole vector soliton. Parameters of the component based on the red valley Hall edge state in Figs. \ref{fig2}(b,c): $k_{y,\alpha}=0.31{\rm K}_y$, $b_\alpha^{\rm nl}=0.0014$ and $b_\alpha^{\prime\prime}=-0.6013$. Parameters of the component based on the blue valley Hall edge state in Figs. \ref{fig2}(b,c): $k_{y,\beta}=0.102{\rm K}_y$, $b_\beta^{\rm nl}=0.003$ and $b_\beta^{\prime\prime}=-1.8188$.
		The velocity of soliton is $v\sim 0.2564$.}
	\label{fig6}
\end{figure}

At the first step, we inspect the propagation of the bright-dipole vector valley Hall edge soliton. For selected Bloch momenta of the components on which we construct vector edge soliton, the first-order derivatives of the edge states are equal to $b'_\nu = -0.1033$. This means that such vector soliton will move in the positive $y$ direction upon propagation. In Fig. \ref{fig5}(a), we display field modulus distributions for vector soliton at different selected propagation distances. Notice that due to its hybrid nature, the constructed soliton is strongly elongated in one direction, along the domain wall. One can see that the soliton propagates with the same velocity, while its profile remains nearly unchanged even after considerable displacements along the domain wall. The dependence of the peak amplitude of the soliton $m_{\rm nlin}=\max\{|\psi|\}$ on propagation distance $z$ in Fig. \ref{fig5}(c) reveals only small oscillations at distances up to $z=10000$. This confirms stability of this composite state involving contribution from two edge states. To illustrate the crucial role of nonlinearity that suppresses diffraction broadening, we checked linear propagation for the same input by removing the nonlinear term in Eq. (\ref{eq1}). Already at $z=2000$ such beam strongly diffracts along the domain wall (but not across it, because it is constructed on topological edge states), as shown in Fig. \ref{fig5}(b). Corresponding peak amplitude $a_{\rm lin}$ displayed in Fig. \ref{fig5}(c) also notably decreases.
We also investigated the propagation of the bright-tripole vector valley Hall edge solitons (see Fig. \ref{fig6}) on the domain wall from Fig. \ref{fig2}(a). The field modulus distributions for such states plotted in Fig. \ref{fig6} clearly reveal considerable contribution from tripole component. Nevertheless, such solitons show stable propagation up to large distances $z$.

\begin{figure}[htbp]
	\centering
	\includegraphics[width=\columnwidth]{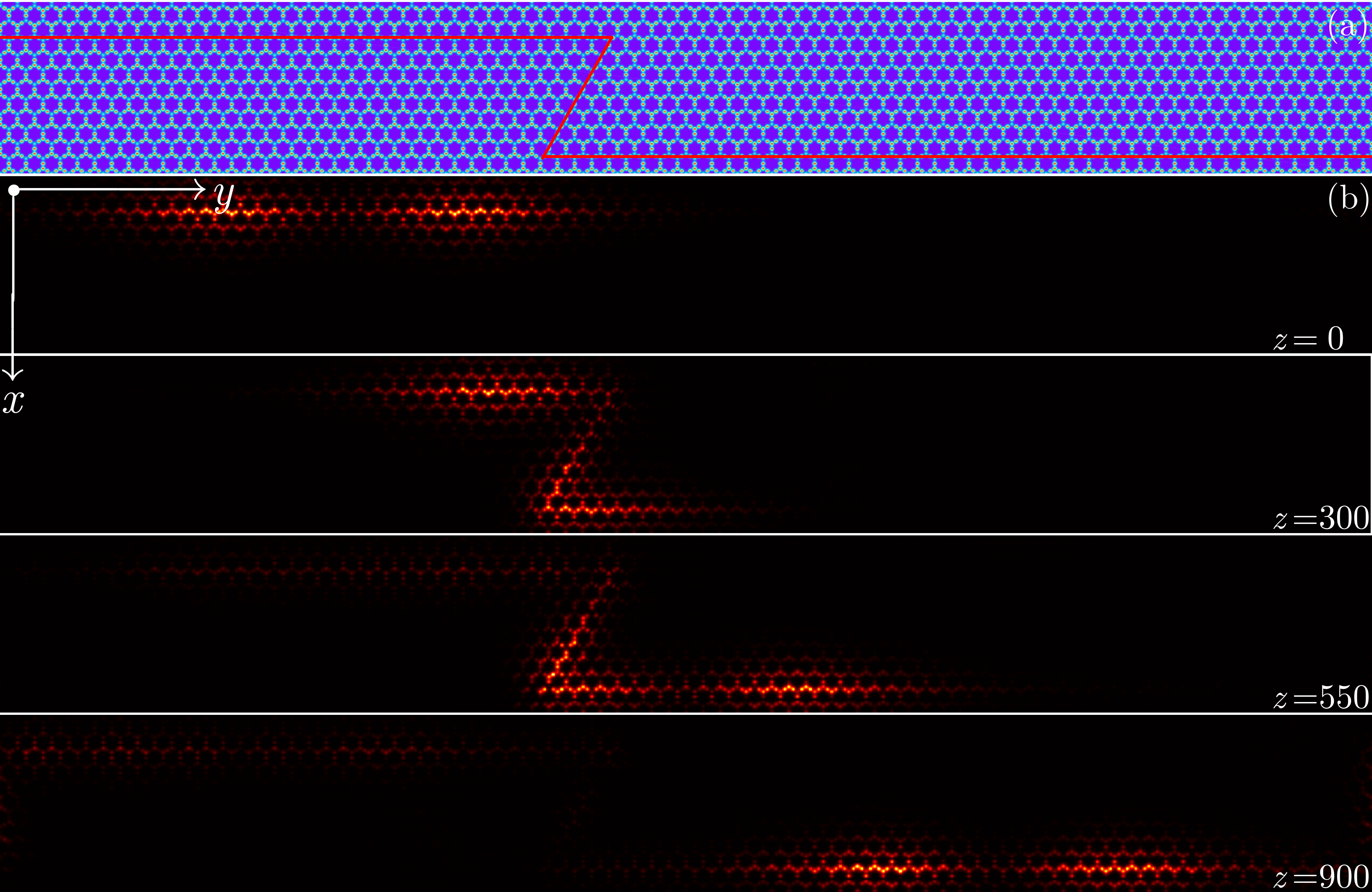}
	\caption{Dynamics of bright-dipole vector soliton passing $Z$-shaped region of the domain wall. Parameters of the component based on the red valley Hall edge state in Figs. \ref{fig2}(b,c): $k_{y,\alpha}=0.31{\rm K}_y$, $b_\alpha^{\rm nl}=0.002$ and $b_\alpha^{\prime\prime}=-0.6013$. Parameters of the component based on the blue valley Hall edge state in Figs. \ref{fig2}(b,c): $k_{y,\beta}=0.102{\rm K}_y$, $b_\beta^{\rm nl}=0.0019$ and $b_\beta^{\prime\prime}=-1.8188$. All panels are shown within the window $-10\le x \le 40$ and $-194\le y\le 194$.}
	\label{fig7}
\end{figure}

Among the representative properties of the valley Hall systems is that topological excitations in them can circumvent sharp corners without backward reflection or radiation into the bulk \cite{tang.oe.29.39755.2021,ren.nano.10.3559.2021}. Here we aim to study whether topological edge solitons show similar level of protection. To this end, we designed a $Z$-shaped domain wall in the superhoneycomb lattice, as marked by the red line in Fig. \ref{fig7}(a). We launched a bright-dipole vector valley Hall edge soliton on the straight segment of the domain wall, so that it propagates towards $Z$-shaped region and recorded field modulus distributions at different distances as illustrated in Fig. \ref{fig7}(b). One can see that soliton follows all bends of the domain wall and circumvents two $60^\circ$ corners without any appreciable backward or bulk reflection. Moreover, the soliton maintains its internal structure after passage of the $Z$-shaped region, as one can clearly see by comparing distributions at $z=0$ and $z=900$. The capability to circumvent sharp corners demonstrates that vector valley Hall edge solitons are indeed topologically protected.

\begin{figure}[htbp]
	\centering
	\includegraphics[width=\columnwidth]{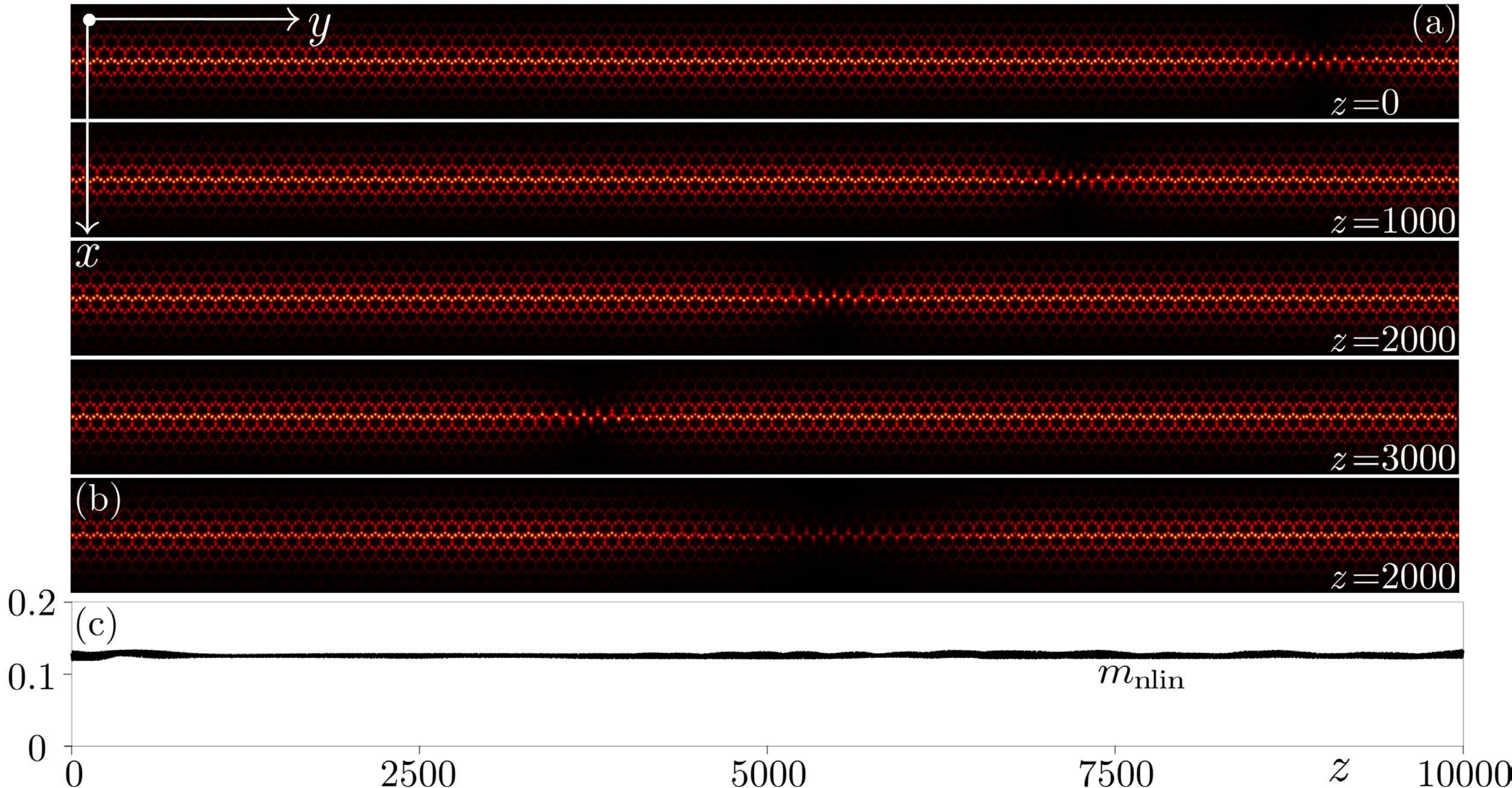}
	\caption{Propagation dynamics of the dark-bright vector solitons. Field modulus distributions at different distances are shown in (a), panel (b) shows result of linear propagation of the same input state as in (a), while peak amplitude during nonlinear propagation is shown in (c). Parameters of the component based on the red valley Hall edge state in Figs. \ref{fig2}(e,f): $k_{y,\alpha}=0.068{\rm K}_y$, $b_\alpha^{\rm nl}=0.004$ and $b_\alpha^{\prime\prime}=0.9236$. Parameters of the component based on the blue valley Hall edge state in Figs. \ref{fig2}(e,f): $k_{y,\beta}=0.04{\rm K}_y$, $b_\beta^{\rm nl}=0.004$ and $b_\beta^{\prime\prime}=1.6209$.
		The velocity of soliton is $v\sim-0.0845$.}
	\label{fig8}
\end{figure}

\begin{figure}[htbp]
	\centering
	\includegraphics[width=\columnwidth]{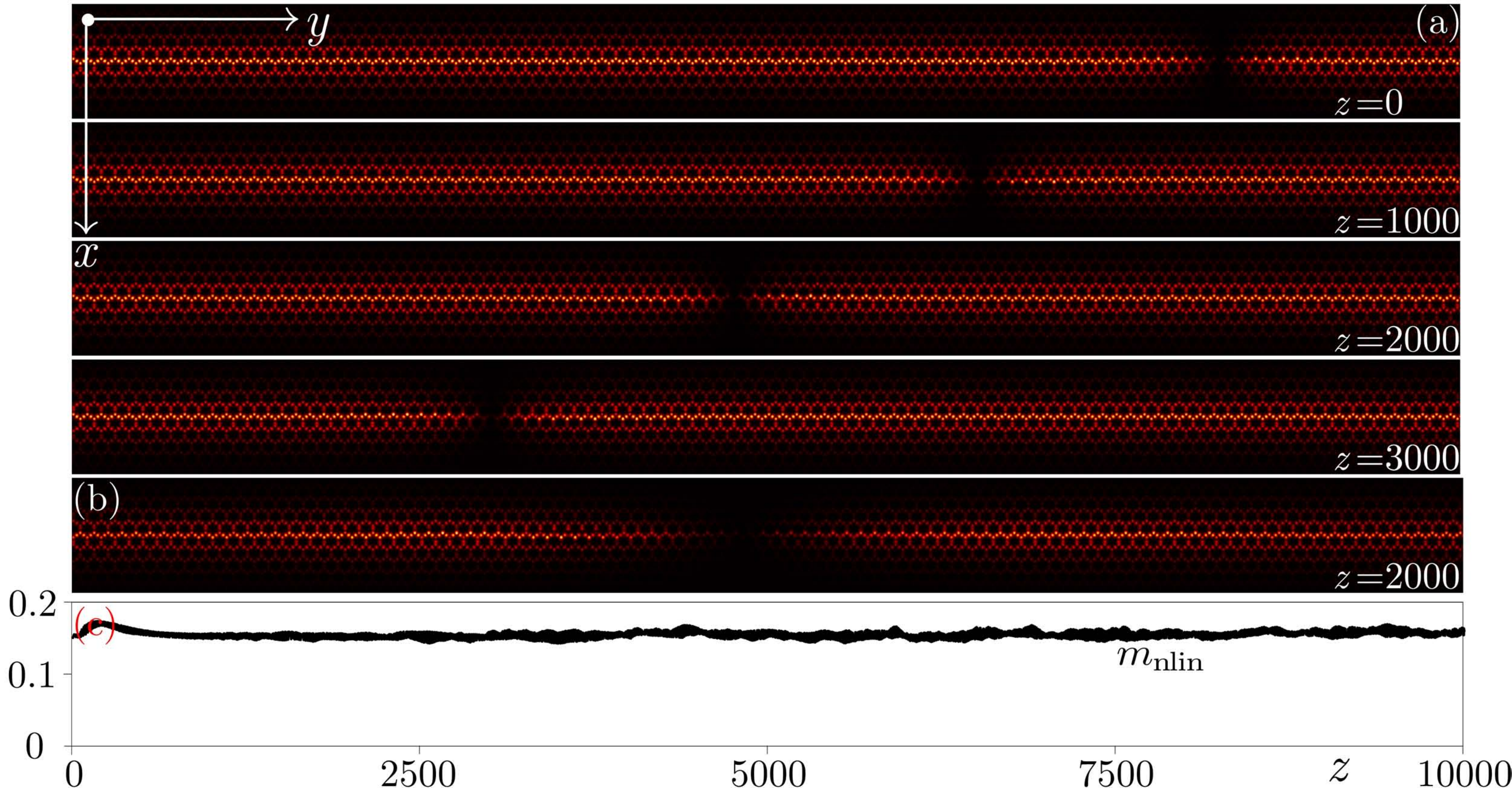}
	\caption{Setup is as Fig. \ref{fig8}, but for the dark-dipole vector soliton. Parameters of the component based on the red valley Hall edge state in Figs. \ref{fig2}(e,f): $k_{y,\alpha}=0.068{\rm K}_y$, $b_\alpha^{\rm nl}=0.0055$ and $b_\alpha^{\prime\prime}=0.9236$. Parameters of the component based on the blue valley Hall edge state in Figs. \ref{fig2}(e,f): $k_{y,\beta}=0.04{\rm K}_y$, $b_\beta^{\rm nl}=0.004$ and $b_\beta^{\prime\prime}=1.6209$.}
	\label{fig9}
\end{figure}

We now turn to the dynamics of the dark-bright and dark-dipole vector solitons at the domain walls depicted in Fig. \ref{fig2}(d). In contrast to bright-dipole and bright-tripole states discussed before, the solitons with one dark component feature nonzero background. The propagation dynamics of the dark-bright vector valley Hall edge soliton is illustrated in Fig. \ref{fig8}, while propagation of the dark-dipole edge soliton is shown in Fig. \ref{fig9}. The field modulus distributions at different distances shown in Figs. \ref{fig8}(a) and \ref{fig9}(a) confirm that such solitons also move with a constant velocity in the negative direction of the $y$-axis, while maintaining their shapes. If nonlinearity is switched off, one immediately observes strong broadening of the dark notch in soliton profile [see Figs. \ref{fig8}(b) and \ref{fig9}(b)]. The amplitude of soliton $m_{\rm nlin}$ (that here characterizes the amplitude of the background) shows only small oscillations in the course of propagation [Figs. \ref{fig8}(c) and \ref{fig9}(c)]. One can also clearly see that background in such solitons remains modulationally stable.

\begin{figure}[htbp]
	\centering
	\includegraphics[width=\columnwidth]{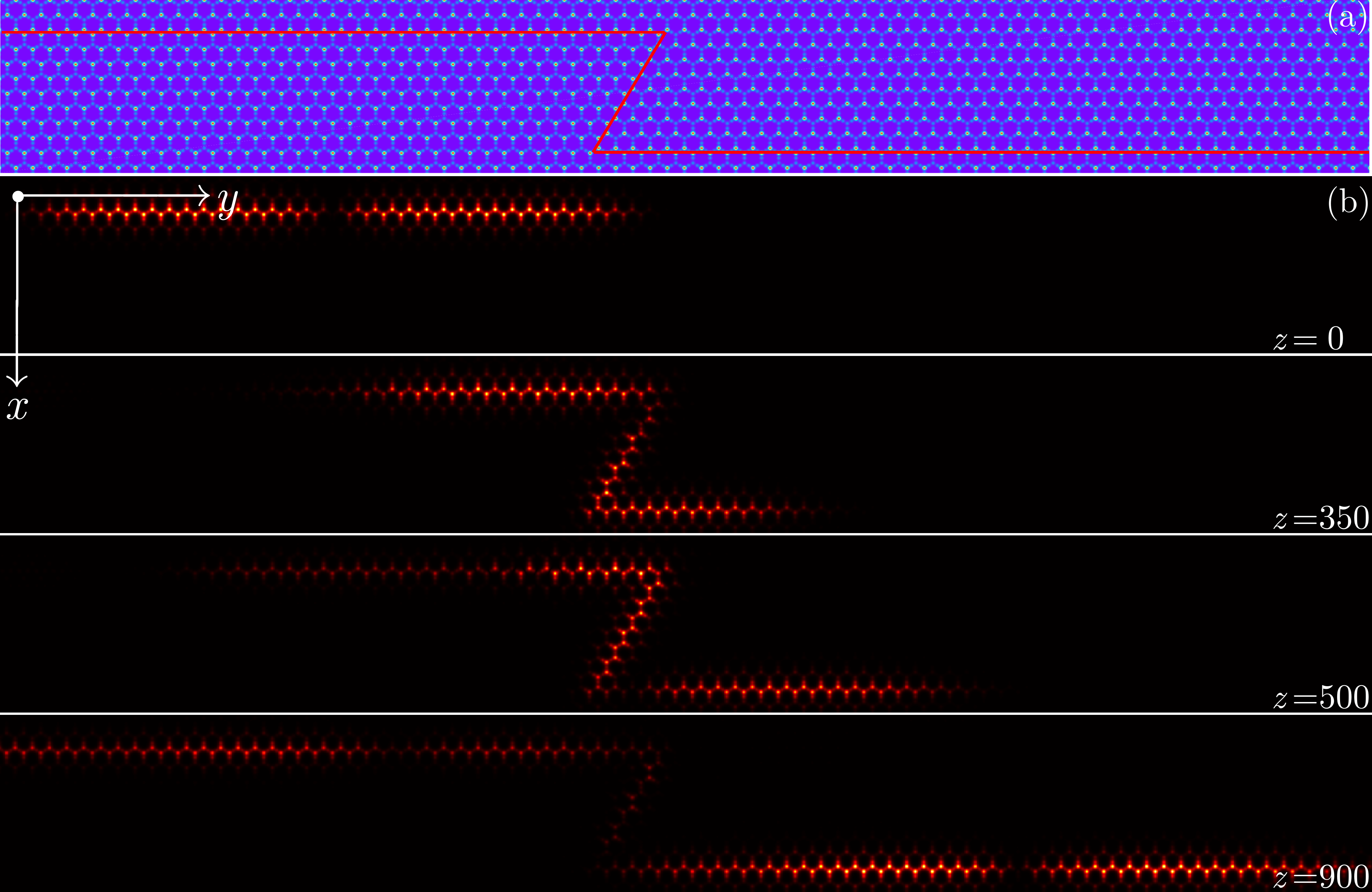}
	\caption{Passage of the scalar dark valley Hall edge soliton through $Z$-shaped domain wall. The structure of the domain wall corresponds to configuration considered in Fig. \ref{fig7}. Parameters for the only component on which such soliton is constructed are $k_y=-0.3{\rm K}_y$, $b_{\rm nl}=0.0015$ and $b^{\prime\prime}=0.2808$.}
	\label{fig10}
\end{figure}

It is necessary to mention that dark topological edge solitons show somewhat weaker protection that depends crucially on the position of the momenta of the edge state on which soliton is constructed with respect to $\bf K$ or $\bf K'$ valleys. Here we verify the degree of protection by studying passage of the scalar dark valley Hall edge soliton via $Z$-shaped domain wall illustrated in Fig. \ref{fig10}. We use here truncated dark soliton to clearly see the outcome of passage. One can see that the notch of the dark soliton remains practically unaffected after passage through $Z$-shaped region, i.e. soliton is not destroyed and follows the domain wall. At the same time, small backward reflection can be seen in Fig. \ref{fig10}(b), which was practically absent for bright states.

\section{Collisions of the valley Hall edge solitons}

Solitonic properties of localized beams ideally imply not only preservation of their shapes upon propagation, but also elasticity of interactions of two such wavepackets. This last aspect of solitonic dynamics has received much less attention than the very existence of solitons. In the case of topological systems, this is partially due to the fact that the velocity of the edge solitons is determined by the group velocity of the respective carrier waves and in the majority of cases such velocity has the same sign along a given branch of the edge states. The use of the valley Hall edge states allows creation of envelope solitons moving in the opposite directions along the edge and, on this reason, it allows to study their interactions upon collisions. In this section we consider several examples of collisions of scalar and vector edge solitons in this system.

Starting with the collision of scalar solitons, we recall that Eq. (\ref{eq5}) with $b''_{\alpha,\beta}\chi_{\alpha,\beta}<0$ possesses a one-component bright soliton solution
\begin{subequations}
\label{eq6}
\begin{align}\label{eq6a}
	A_{\alpha,\beta} = &\sqrt{\frac{2 b_{\alpha,\beta}^{\rm nl} }{ \chi_{\alpha,\beta} }} {\rm sech} \left[ \sqrt{\frac{-2 b_{\alpha,\beta}^{\rm nl}}{b''_{\alpha,\beta}}}  (x-v_{\alpha,\beta}t) \right] e^{  ib_{\alpha,\beta}^{\rm nl} z },
	\\
	\label{eq6b}
	A_{\beta,\alpha}=&0
\end{align}
\end{subequations}
 Notice that unlike in the ansatz (\ref{eq4}) exploring a common frame defined by the running variable $Y$, one formally cannot introduce such a frame for two solitons with $v_\alpha\neq v_\beta$. Since the group velocities of the carrier modes define the velocities of solitons in the laboratory frame, the envelopes given by Eq.~(\ref{eq6}) must be considered in different frames. Thus, being far from each other such solitons are described by the nonlinear Schr\"odinger equations in different frames. Such description fails during the interaction time, when the envelopes strongly overlap, and thus, experience nonlinear interactions (i.e., cannot be considered separately). In the interaction interval analytical descriptions based on slowly varying amplitudes is not applicable anymore, and the dynamics must be studied numerically.

\begin{figure*}[htbp]
	\centering
	\includegraphics[width=\textwidth]{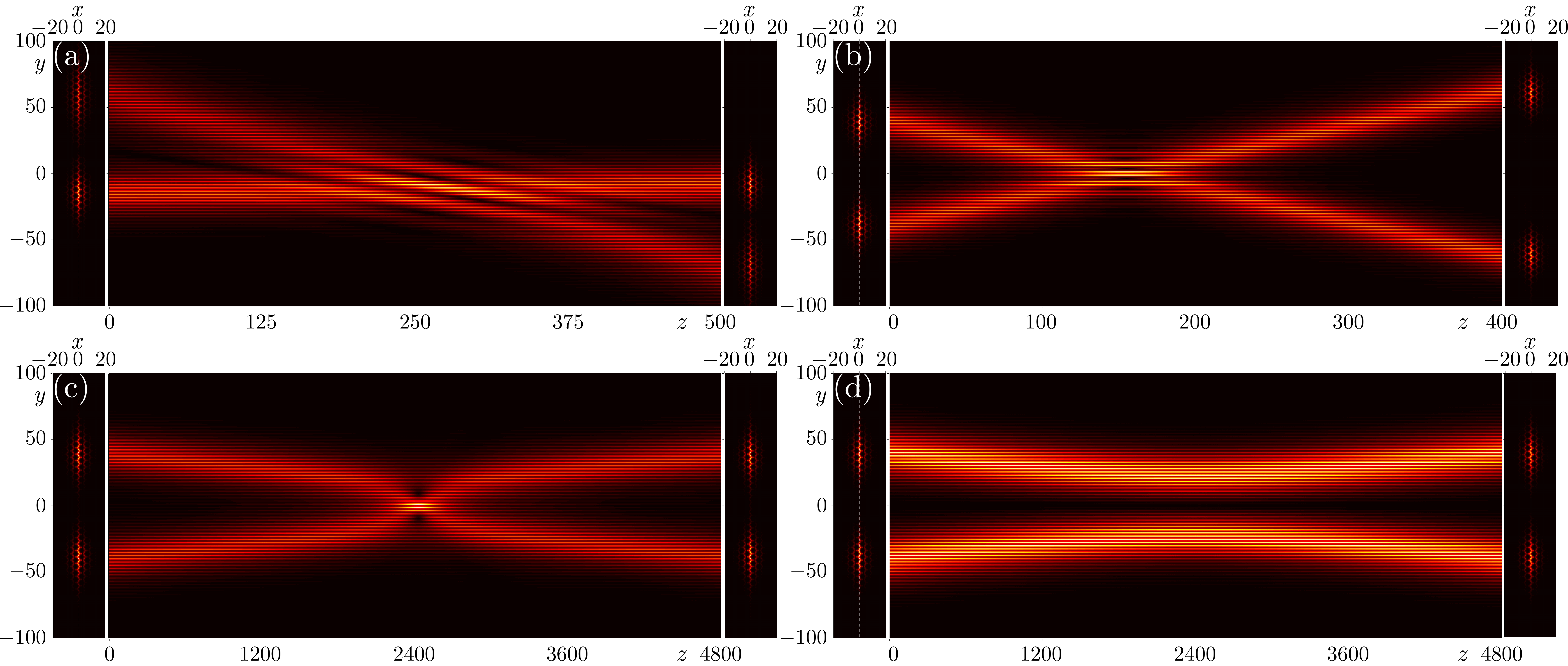}
	\caption{Frontal collision of two bright valley Hall edge solitons $\psi_1$ and $\psi_2$ bifurcating from the red valley Hall edge states in Fig. \ref{fig2}(b). (a) Left panel: two input solitons with the top soliton being $\psi_1$ and the bottom one being $\psi_2$. Dashed line shows the cross-section at $x=0$ along which we show collision dynamics. Middle panel: collision dynamics in the cross section. Right panel: output soliton profiles. Parameters for $\psi_1$ are: $k_y=-0.3{\rm K}_y$, $b'=0.2487$, $b''=-0.605$, and $b_{\rm nl}=0.002$. Parameters for $\psi_2$ are: $k_y=0.01{\rm K}_y$, $b'=-0.0085$, $b''=-0.6567$, and $b_{\rm nl}=0.005$. (b) The same as in (a), but solitons bifurcate from the edge states at $k_y=\pm0.3{\rm K}_y$, with $b_{\rm nl}=0.005$ for both solitons, and they have opposite by sign and equal by modulus propagation velocities. (c) The same as in (b), but solitons bifurcate from the edge states at $k_y=\pm 0.01{\rm K}_y$. (d) The same as in (c), but with a relative initial $\pi$-phase shift between two solitons.
	}
	\label{fig11}
\end{figure*}

We first simulate the collision of scalar bright solitons bifurcating from two different points of the red valley Hall edge branch in Fig.~\ref{fig2}(b) using the envelope Eq.~(\ref{eq6}) to generate the input solitons. The respective states will be denoted by $\psi_1$ and $\psi_2$.
Collision dynamics was simulated using fully two-dimensional Eq. (\ref{eq1}). Concentrating on the frontal collisions, we consider two solitons with different (by modulus) velocities in Fig.~\ref{fig11}(a), while collision of solitons with equal by modulus, but opposite by sign velocities is shown in Fig.~\ref{fig11}(b).
In each subfigure (a)-(d) the left outermost panel shows two-dimensional input field distribution, while right outermost panel shows the output field distribution. To illustrate the dynamics of the collision, in the middle panels in each subfigure we show the evolution of the field modulus $|\psi|$ in the cross section along the domain wall (indicated by the dashed while line on the left panels). In both subfigures (a) and (b) we observe nearly elastic collision of solitons. The total amplitude of the field increases in the collision regions. After collision the solitons keep moving with nearly the same group velocities as before collision. It is relevant to note that when the relative $\pi$-phase shift is introduced into one of the colliding solitons, the interference fringes in the interaction region change considerably, but the output field modulus distributions observed well beyond collision point remain practically unaffected.

The frontal collision dynamics changes dramatically if the solitons have sufficiently small initial velocities, corresponding to $k_y=\pm0.01{\rm K}_y$, as illustrated in Figs. \ref{fig11}(c,d). Such collision can be formally described by the scalar limit of the model (\ref{eq5}) with zero carrier-wave group velocity $v=0$, but with nonzero initial velocities of the colliding solitons: a sufficiently small group velocity can be accounted by the second order of the multiple-scale expansion. As a result, such collision becomes strongly dependent on the phases of solitons. In close similarity to the interactions of conventional solitons described by the nonlinear Schr\"odinger equation~\cite{aitchison.ol.16.15.1991, stegeman.science.286.1518.1999}, initially in-phase solitons pass through each other acquiring relative shifts of the output trajectory with respect to the input one. One can observe this by comparing the linear trajectories before and after the collision in Fig.~\ref{fig11}(c). The out-of-phase solitons repel each other as clearly seen in Fig.~\ref{fig11}(d). It should be stressed that even for small collision velocities no radiation emission is observed in the collision process.

\begin{figure*}[htbp]
	\centering
	\includegraphics[width=\textwidth]{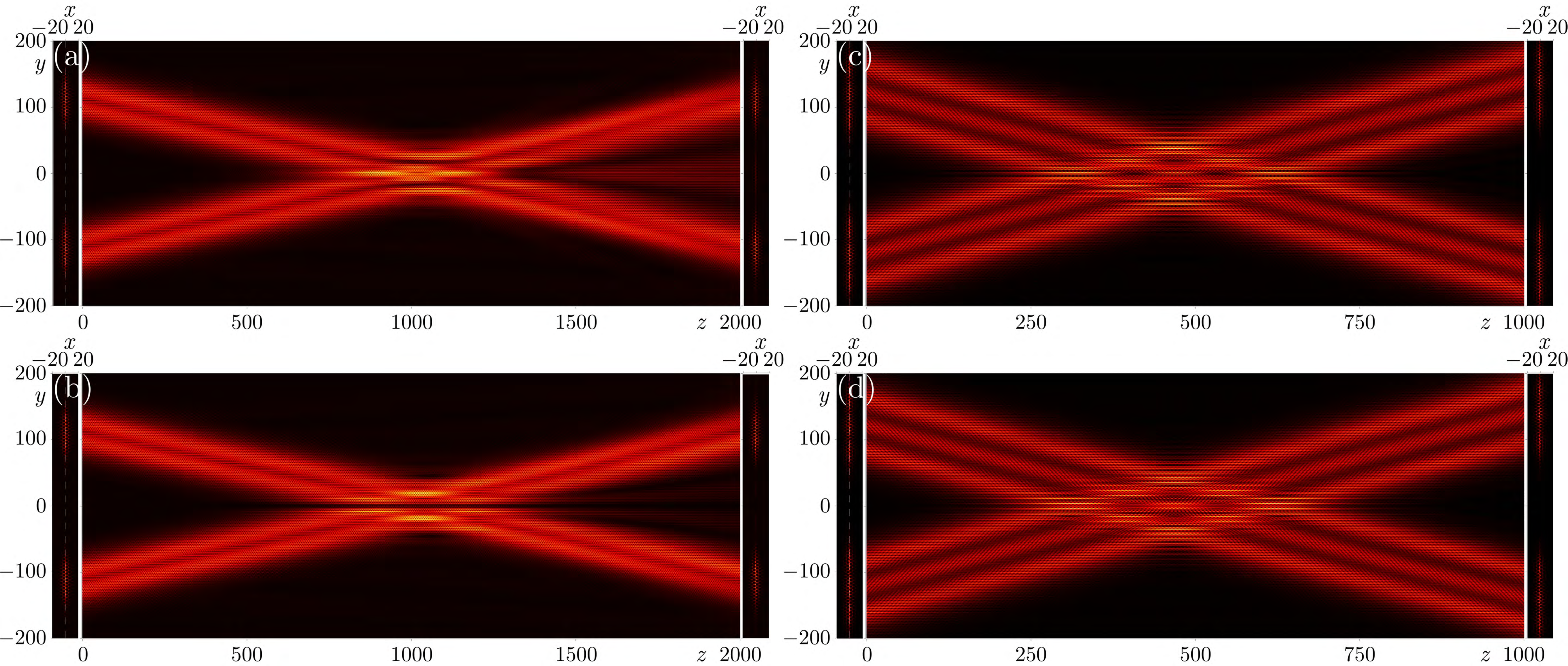}
	\caption{(a) Collision of two vector bright-dipole valley Hall edge solitons $\psi_1$ (top) and $\psi_2$ (bottom).
		Parameters for $\psi_2$ are same as those adopted in Fig. \ref{fig5}.
		Bloch momenta and $b'$ adopted for $\psi_1$ are opposite to those for $\psi_2$, but other parameters are the same.
		(b) Setup is as (a),
		but there is a $\pi$-phase shift between two solitons.
		(c,d) Setup is as (a,b) but for two vector bright-tripole valley Hall edge solitons.
		The input parameters for $\psi_2$ are same as in Fig. \ref{fig6}.
	}
	\label{fig12}
\end{figure*}

For the sake of completeness, in Figs. \ref{fig12}(a) and \ref{fig12}(b), we show collisions of two in-phase and out-of-phase vector bright-dipole solitons, respectively, with the same parameters as in Fig.~\ref{fig5}.
One observes that such vector solitons maintain their profiles and internal structure after the collision, as one can see from the structure of intensity dips in soliton profile. However, one can also observe that radiative losses are stronger upon collisions of vector solitons, whose envelopes are described by the nonintegrable model in Eq. (\ref{eq5}) [see Figs.~\ref{fig12}(a,b)]. Radiation typically appears between vector edge solitons after the collision event and it always remains at the domain wall. These radiative losses can be considerably suppressed by increasing the relative velocities of the colliding edge solitons, as shown in Figs. \ref{fig12}(c,d) for vector bright-tripole solitons with the same parameters as those adopted in Fig. \ref{fig6}. Thus, solitons in Figs. \ref{fig12}(c,d) have velocities $|v| \sim 0.2564$ in contrast to solitons from Figs. \ref{fig12}(a,b) with $|v| \sim 0.1033$. No obvious radiation is visible after collision of such bright-tripole states indicating on practically elastic collision regime at sufficiently high velocities.

\section{Conclusions}

Summarizing, we have reported on the existence of a rich variety of topological vector edge solitons in valley Hall system, at the domain wall between two superhoneycomb lattices. Such solitons are constructed as envelope solitons on two coexisting topological edge states from different gaps and with different Bloch momenta. The states that we are using for construction of the envelope solitons not only provide equal signs of the effective dispersion coefficients, but can also be selected to have equal group velocities, that is a necessary requirement for the formation of vector solitons, whose components remain always bound to each other in the course of propagation. The families of solitons that we obtained include bright-dipole, bright-tripole, dark-bright, and dark-dipole vector solitons. Despite their complex internal structure and rapid displacement along the domain, such states demonstrate remarkable robustness upon propagation and maintain their internal structure even upon passage of the sharp bends of the domain wall. This illustrates topological nature and protection of the respective solutions. Valley Hall system proposed here allowed us to study collisions of both scalar and vector edge solitons. It was found that collisions of scalar solitons are nearly elastic, while collisions of vector solitons are elastic for large velocities, but can be accompanied by the emission of radiation for small collision velocities (radiation nevertheless remains at the domain wall). Our results illustrate considerable potential of valley Hall topological systems for realization of previously unknown nonlinear topological states and phenomena.

\section*{Funding}
National Natural Science Foundation of China (12074308, U1537210);
Russian Science Foundation (21-12-00096);
Portuguese Foundation for Science and Technology (FCT) (UIDB/00618/2020).

%
%


\end{document}